\documentstyle[12pt,preprint,aps,floats,epsf]{revtex}
\def\msun{M$_{\odot}$}
\def\ni{\noindent}
\begin{document}
\title{Quark-Hadron Phase Transitions in Young and Old Neutron Stars}
\author{A. Steiner, M. Prakash, and J.M. Lattimer}
\address{
Department of Physics \& Astronomy, SUNY at Stony Brook, New York
11794-3800}
\date{\today}
\maketitle
\begin{abstract}

The mixed phase of quarks and hadrons which might exist in the dense
matter encountered in the varying conditions of temperature and
trapped neutrino fraction in proto-neutron stars is studied.  The
extent that the mixed phase depends upon the thermodynamical
parameters as well as on the stiffness of matter in the hadronic and
quark phases is discussed.  We show that hadronic equations of state
that maximize the quark content of matter at a given {\it density}
generally minimize the extent of the mixed phase region in a neutron
star of a given mass, and that only in extreme cases could a pure
quark star result.  For both the Nambu Jona-Lasinio and MIT bag quark
models, neutrino trapping inhibits the appearance of a mixed phase
which leads to possible proto-neutron star metastability.  The main
difference between the two quark models is the small abundance of
strange quarks in the former. We also demonstrate that $\partial
T/\partial n<0$ along adiabats in the quark-hadron mixed phase,
opposite to what is found for the kaon condensates-hadron mixed phase.
This could lead to core temperatures which are significantly lower in
stars containing  quarks than in those not containing
quarks.

\bigskip\noindent
PACS: 97.60.Jd, 21.65.+f, 12.39.Fe, 26.60.+c   \\

\end{abstract}

\newpage


It has been proposed by several authors \cite{Glen,HPS,PCL} that a
mixed phase of hadrons and deconfined quarks might exist in the high
density interior of neutron stars.  The quark-hadron transition is
being probed in relativistic heavy-ion collisions \cite{QM99}, but
observations of neutron stars, in which matter is considerably less
energetic, might also provide evidence of its existence. In this
work, we analyze the structure of young and old neutron stars that
contain quarks and indicate what effect quarks might make on
observations, in particular, observations of neutrino emissions.

When a neutron star is born, the neutrinos produced by electron
capture in the beta-equilibrated matter are prevented by their short
mean free paths from leaving the star on dynamical timescales. The
number of leptons per baryon that remain trapped is approximately 0.4,
the precise value depending on the efficiency of electron capture
reactions during the gravitational collapse of the progenitor star.
On a timescale of 10--20 seconds, the neutrinos diffuse from the star,
but leave behind much of their energy which causes significant heating
of the ambient matter~\cite{BML,BL}.  Entropies per baryon of about 2
(in units of the Boltzmann constant $k_B$), and temperatures in the
range 30--50 MeV, are generally achieved in the inner 50\% of the
star's core at the peak of the heating.  This is to be compared to
entropies of approximately 1 which exist in the initial configuration.  
Following the heating, the star cools by radiating
neutrino pairs of all flavors, and temperatures fall to below 1 MeV
within minutes.  Recent calculations~\cite{KJ,Pons99} have verified
this general scenario for a variety of equations of state (EOS) and
assumptions about the composition of high-density matter.

Compared to cold neutron stars, the appearance of quarks is suppressed
in a proto-neutron star (PNS) because of its high lepton number
content.  As the neutrinos leak out of a PNS, however,
the central density of the star increases and the threshold density
for the appearance of quarks decreases. Previous studies \cite{PCL}
have shown that the maximum mass supported by neutrino-rich matter is
larger than that supported by neutrinoless matter if quarks appear.
This gives rise to the possibility that some PNSs might
become metastable~\cite{TPL,prep,SN87}, which would occur if the
PNS mass lies within this range of maximum masses.
When the maximum mass decreases below the PNS mass,
after most of the neutrinos have diffused from the star, a collapse to
a black hole ensues.

It is an open question whether or not the appearance of quarks could
produce an observable effect in the light curves of the emitted
neutrinos.  To date, studies of quarks in PNS evoluton
have ignored finite temperature, and have been limited mostly to the MIT bag
model of quarks and restricted models of hadronic interactions.  The new
features of our study are to 1) include both the effects of trapped
neutrinos and finite temperature, 2) examine the role of the quark
model by employing both the traditional MIT bag model and the
Nambu Jona-Lasinio (NJL) quark Lagrangian, 3) explore the effects of
stiffness of the hadronic interactions on the quark-hadron transition,
4) study the effect of hyperons, and 5) delineate the phase diagram
in the lepton number--baryon number density plane, appropriate for
PNS studies.


To model the hadronic phase, we use a field-theoretical description,
in which baryons interact via the exchange of $\sigma$-, $\omega$-,
and $\rho$- mesons, extended to include hyperons. Specifically, we
follow the approach of M\"uller and Serot \cite{msl} (hereafter
MS). There is limited data which constrains the hadronic EOS at
densities between nuclear matter equilibrium density and the density
at which quarks become deconfined.  Based on the considerations of
``naturalness'' in the context of an effective field theoretical
approach, the Lagrangian gives a range of possible values for the
couplings of higher order interactions between the vector mesons of
the theory. We explore a range of these couplings in order to test the
sensitivity of our results to variations in the stiffness of the
hadronic supernuclear 
EOS.  For densities lower than 0.08 $\mathrm{fm^{-3}}$, we
use the zero-temperature EOS of Negele and Vautherin \cite{Negele},
and for densities lower than 0.001 $\mathrm{fm^{-3}}$ we use the
Baym-Pethick-Sutherland zero-temperature EOS \cite{bps}.  Since the
maximum mass and central densities of neutron stars depend only
marginally on the low-density EOS, the assumption of zero-temperature
for this low-density matter is satisfactory.

The MS Lagrangian is
\begin{eqnarray}
{\cal L} &=& \sum_{B} \bar{B} \left(i\gamma^{\mu}\partial_{\mu}-
g_{\omega B} \gamma^{\mu} \omega_{\mu} -
g_{\rho B} \gamma^{\mu} {\bf b}_{\mu} \cdot {\bf t} - M_B +
g_{\sigma B} \sigma \right) B  \nonumber \\
&-& \frac{1}{2}m^2_{\sigma} \sigma^2
+\frac{1}{2}\partial_{\mu}\sigma\partial^{\mu}\sigma -
\frac{\kappa}{3!}\sigma^3 - \frac{\lambda}{4}\sigma^4  \nonumber \\
&+& \frac{1}{2}m^2_{\omega}\omega^{\mu}\omega_{\mu} -
\frac{1}{4}F_{\mu \nu} F^{\mu \nu}
+ \frac{\zeta}{4!}g_{\omega}^4\left(\omega^\mu \omega_\mu\right)^2
\nonumber \\
&+& \frac{1}{2}m^2_{\rho} {\bf b}^{\mu} {\bf b}_{\mu} -
\frac{1}{4}{\bf B}_{\mu \nu} {\bf B}^{\mu \nu}
+ \frac{\xi}{4!}g_{\rho}^4\left({\bf b}^\mu {\bf b}_\mu\right)^2 \,
+ {\cal L}_\ell \,
\end{eqnarray}
where the sum over $B$ is a sum over all nucleons and hyperons, and
${\cal L}_\ell$ represents the sum of the Dirac Lagrangians for all of
the leptons (electrons, muons and neutrinos). The values of $\kappa$,
$\lambda$, $g_{\rho N}$, $g_{\sigma N}$, and $g_{\omega N}$ are set by
matching the equilibrium nuclear density ($n_0=0.16~
\mathrm{fm}^{-3}$), binding energy ($E_b=-16$ MeV), compressibility
($K_0 =250$ MeV), nucleon effective mass ($M^{*}_0 = 0.6 M$), and
symmetry energy ($e_{sym}=35$ MeV) at $n_0$. The remaining
two parameters, $\xi$ and $\zeta$, associated with non-linear vector
and isovector interactions, control the stiffness of the hadronic EOS
at supernuclear densities. Larger values of either parameter tend to
soften the EOS. The acceptable ranges for $\zeta$ and $\xi$, based on
considerations of naturalness, are $0 \leq \xi \leq 1.5$ and $0
\leq \zeta \leq 0.06$ \cite{msl}.

We include the $\Lambda$, $\Sigma^{+}$, $\Sigma^{0}$, $\Sigma^{-}$,
$\Xi^{0}$, and $\Xi^{-}$ hyperons and ignore the heavier $\Delta$
baryon which is too massive to affect our results. We assume that all six
hyperon coupling constants with a particular vector meson are equal.  
Furthermore, the hyperon coupling constants are related to the 
nucleon--vector meson coupling contants by
\begin{equation}
g_{\sigma H} = x_{\sigma} ~ g_{\sigma N} \,,\qquad
g_{\rho H} = x_{\rho} ~ g_{\rho N} \,,\qquad
g_{\omega H} = x_{\omega} ~ g_{\omega N}\,.
\end{equation}
Following Glendenning and Moszkowski \cite{Glen2} we assume
$x_\rho=x_\sigma=0.8$.  We also take $x_\omega=0.895$, which  follows
from the binding energy, $-28$ MeV, of the $\Lambda$ hyperon in nuclei
\cite{Millener}.

In the mean field approximation, the thermodynamic potential $\Omega$
is given by
({\it cf.} \cite{Knorren})
\begin{eqnarray}
\frac {\Omega}{V} = &&
\frac{1}{2}m^2_{\sigma} \sigma^2
+ \frac{\kappa}{3!}\sigma^3 + \frac{\lambda}{4!}\sigma^4
- \frac{1}{2}m^2_{\omega}\omega_0^2
- \frac{1}{2}m^2_{\rho} b_0^2
- \frac{\zeta}{4!}g_{\omega}^4 \omega_0^4 \nonumber \\
&& - \frac{\xi}{4!}g_{\rho}^4 b_0^4
- \sum_B{2 T \int{\frac{d^3 p}{\left(2 \pi \right)^3}
\ln{\left(1-f_B\right)}}} + \Omega_\ell \,
\end{eqnarray}
where the distribution function is 
$f_B =[1 + \exp(\beta(E^{*}_B-\nu_B))]^{-1}$.
Here, $\beta=1/T$, the effective chemical potential is 
$ \nu_B  = \mu_B  - g_{\omega B} \omega_0 - g_{\rho B} t_3 b_0 $,
the effective mass is $ M ^{*}_B = M_B  - g_{\sigma B}\sigma$, and
$E^{*}_B=\sqrt{p^2+M_B^{* 2}}.$  The contribution of antibaryons 
is not significant for the thermodynamics of interest for a PNS and have been 
ignored.  
The contribution from the leptons,  
$\Omega_\ell$, is given adequately by its non-interacting form \cite{Kapusta}. 


The thermodynamic potential of the quark phase is $\Omega =\Omega_{\mathrm{FG}}
+ \Omega_{\mathrm{Int}}$, where
\begin{eqnarray}
\frac{\Omega_{\mathrm{FG}}}{V} = - 2 N_c T \sum_{i=u,d,s}
\int \frac{d^3 p}{\left(2 \pi\right)^3}
\left[ \ln{(1-f_i)} + \ln{(1-{\bar f_i})} \right]\,
\label{FG}\end{eqnarray}
denotes the Fermi gas contribution arising from quarks.  We
consider three flavors, $i=u,d,s$ and three colors, $N_c=3$ of quarks.
The distribution functions of fermions and anti-fermions are $f_i=[1 +
\exp(\beta(E_i-\mu_i))]^{-1}$ and ${\bar f_i} = [1 +
\exp(\beta(E_i+\mu_i))]^{-1}$, where $E_i$ and $\mu_i$ are the single
particle energy and chemical potential, respectively, of quark species
$i$.  To explore the sensitivity of the quark model, we contrast the
results of the MIT bag and the Nambu Jona-Lasinio (henceforth NJL)
models for $\Omega_{\mathrm{Int}}$.

In the MIT bag model, the Fermi gas contribution is calculated using
current, as opposed to  
dynamical, quark masses. The interactions between quarks
inside the confining cavity (the bag) are taken to be
perturbative. Thus, $\Omega_{\mathrm{Int}} = BV + \Omega_{ex} +
\Omega_{corr} + \cdots$, where the constant $B$ has the simple
interpretation as the pressure of the vacuum (the
so-called bag constant or bag pressure), $\Omega_{ex}$ denotes the
two-loop or one-gluon exchange contribution, and $\Omega_{corr}$
represents higher order correlation contributions from ring diagrams,
etc. \cite{Kapusta} In this work, we will restrict ourselves to the
simplest bag model and keep only the constant cavity pressure term.
The results are qualitatively similar to what is obtained by including
the higher order terms, if the bag constant $B$ is slightly altered
\cite{PBP}.

Several features of the Lagrangian of Quantum Chromo-Dynamics 
(QCD), including the spontaneous breakdown of chiral 
symmetry, are exhibited by the Nambu Jona-Lasinio (NJL) model, which
shares many symmetries with QCD.  In its commonly used form, the NJL
Lagrangian reads
\begin{eqnarray}
{\cal L} &=& \bar q ( i \partial{\hskip-2.0mm}/ - {\hat m_0}) q \;+\;
G \sum_{k=0}^8 [\,({\bar q}\lambda_k q)^2 + ({\bar q}
i\gamma_5\lambda_k q)^2\,] \nonumber \\ &-& K \,[ \,{\rm det}_f ({\bar
q}(1+\gamma_5) q) + {\rm det}_f ({\bar q}(1-\gamma_5) q) \,] \ . \label{L3}
\end{eqnarray}
The determinant operates over flavor space, ${\hat m_0}$ is the 3
$\times$ 3 diagonal current quark mass matrix, $\lambda_k$ represents
the 8 generators of SU(3), and $\lambda_0$ is proportional to the
identity matrix.  The four-fermion interactions stem from the original
formulation of this model \cite{NJL}, while the flavor mixing,
determinental interaction is added to break $U_A(1)$ symmetry
\cite{tHooft}.  
Since 
the coupling constants $G$ and $K$ are dimensionful,
the quantum theory is non-renormalizable. 
Therefore, an ultraviolet
cutoff $\Lambda$ is imposed, and results are considered meaningful
only if the quark Fermi momenta are well below this cutoff.

The coupling constants $G$ and $K$, the strange quark mass $m_{s,0}$,
and the three-momentum ultraviolet cutoff parameter $\Lambda$, are
fixed by fitting the experimental values of $f_\pi$, $m_\pi$, $m_K$
and $m_{\eta'}$.  We use the values of Ref.~\cite{Rehberg}, namely
$\Lambda = 602.3$ MeV, $G\Lambda^2 = 1.835$, $K\Lambda^5 = 12.36$, and
$m_{0,s}=140.7$ MeV, obtained using $m_{0,u}=m_{0,d}=5.5$ MeV.  The
subscript ``$0$'' denotes current quark masses.   Results 
of the gross properties of PNSs  
obained by the alternative fits of Refs. \cite{parms2} and
\cite{Hatsuda} are similar to the results quoted below. 

In the mean field approximation at finite temperature and at finite baryon 
density, the
thermodynamic potential due to interactions is given by \cite{Hatsuda}:
\begin{eqnarray}
\frac{\Omega_{\mathrm{Int}}}{V} &=& 2 N_c \sum_{i=u,d,s}
\int \frac {d^3p}{(2\pi^3}
\left( {\sqrt{m_i^2 + p^2}} - {\sqrt{m_{0,i}^2 + p^2}} \right) \nonumber \\ 
&+& 2 G \langle\bar{q}_i q_i \rangle^2
- 4 K \langle \bar{q}_u q_u \rangle \langle \bar{q}_d q_d \rangle
\langle \bar{q}_s q_s \rangle\,.
\label{omegint}
\end{eqnarray}
In both Eqs. (\ref{FG}) and (\ref{omegint}) for the NJL model,
the quark masses are dynamically generated
as solutions
of the gap equation obtained by requiring that the potential be stationary with
respect to variations in the quark condensate $\langle {\bar{q}_i q_i}\rangle$:
\begin{equation}
   m_i = m_{0,i} - 4 G \langle {\bar{q}_i q_i}\rangle +
     2 K \, \langle{\bar{q}_j q_j}\rangle \langle{\bar{q}_k q_k}\rangle \ ,
\label{gap}
\end{equation}
$(q_i,q_j,q_k)$ representing any permutation of $(u,d,s)$.
The quark condensate
$\langle {\bar{q}_i q_i}\rangle$ and the quark number density
$n_i=\langle {q_i^{\dagger} q_i}\rangle$ are given by:
\begin{eqnarray}
\langle{\bar{q}_i q_i}\rangle & = & - 2 N_c  \int
{ \frac{d^3 p}{\left(2 \pi \right)^3} \frac {m_i}{E_i} 
\left[1-f_i-{\bar f_i}\right]  }  \nonumber \\
n_i=\langle {q^{\dagger}_i q_i}\rangle & = & 2 N_c \int
{ \frac{d^3 p}{\left(2 \pi \right)^3} \left[f_i-{\bar f_i}\right]}
\,.
\end{eqnarray}

A comparison between the MIT bag and NJL models is facilitated by
defining an effective bag pressure in the NJL model to be \cite{Buballa}
$B_{eff}=\Omega_{\mathrm{int}}/V-B_0$ with $B_0 V =
\Omega_{\mathrm{int}}|_{n_u=n_d=n_s=0}$ a constant value which makes
the vacuum energy density zero.  In this way, the thermodynamic
potential can be expressed as $\Omega= B_{eff}V + \Omega_{\rm FG}$ which
is to be compared to the MIT bag result $\Omega=BV+\Omega_{\rm
FG}$.  Note, however, that $\Omega_{FG}$ in the NJL model is
calculated using the dynamical quark masses from Eq.~(\ref{gap}).


Both PNS and neutron star matter are in beta equilibrium, which
together with charge conservation implies
\begin{eqnarray}
\mu_e - \mu_{\nu_e} = \mu_{\mu} - \mu_{\nu_{\mu}} \,;\qquad
\mu_B = b_i \mu_n - q_i \mu_e + q_i \mu_{\nu_e}\,,
\end{eqnarray}
where $b_i$ and $q_i$ are the baryon number and charge, respectively,
of the hadron or quark species $i$.  We ignore surface and Coulomb
effects for the structure in the mixed phase so the leptons are
everywhere free Fermi gases.

The initial PNS contains trapped neutrinos, so the
electron and muon lepton numbers may be assumed fixed:
\begin{eqnarray} 
Y_{L_e} \equiv
\frac {n_e+n_{\nu_e}} {n_B} = 0.4\,; \qquad Y_{L_{\mu}} \equiv \frac
{n_{\mu}+ n_{\nu_{\mu}}}{n_B} = 0\,. 
\end{eqnarray} 
Also, calculations generally show that the entropy of the inner half
of the star has an entropy per baryon $s\approx1$.  In the cases in
which the neutrinos have completely escaped, the neutrino chemical
potentials and densities are vanishingly small.  The departing
neutrinos maximally heat the stellar interior to entropies around 2
per baryon after approximately 10--20 seconds\cite{BL,KJ,Pons99}.
After several minutes, neutrino cooling reduces the temperature to
essentially zero on the scale of MeVs.  Thus, we consider three
approximate entropies and compositions to represent the thermodynamic
conditions in an evolving PNS: the initial state ($s=1,Y_{L_e}=0.4$),
the maximally heated star ($s=2, Y_{\nu_e}=0$), and the cold,
catalyzed star ($s=0, Y_{\nu_e}=0$).  Of course, treating the PNS as a
monolithic structure of fixed entropy and composition is an
oversimplification, and full evolutionary calculations are required to
confirm these estimates.

Quarks are assumed to appear by forming a mixed phase with the hadrons
satisfying Gibbs' rules for phase equilibrium.  Matter in this mixed
phase is in thermal, mechanical and chemical equilbrium, so that
\begin{eqnarray}
P^I &=& P^{II}\,; \qquad    \mu_n = 2\mu_d + \mu_u\,, 
\end{eqnarray}
where $I$ and $II$ denote the hadronic and quark phases, respectively. The
restriction that the pure phases I and II are independently charge
neutral is replaced by the condition of global charge neutrality \cite{Glen}
\begin{eqnarray} 
\chi n^I_c + (1- \chi) n^{II}_c = 0 \,, 
\end{eqnarray}
where $n_c$ is the charge density and $\chi$ is the volume fraction of
the hadronic phase.  The energy and entropy densities in the mixed
phase can be expressed in terms of the corresponding quantities
in the hadronic and quark phases:
\begin{eqnarray} 
\varepsilon = \chi
\varepsilon^I + \left(1-\chi\right) \varepsilon^{II} \ , \qquad s = \chi s^I +
\left(1-\chi\right) s^{II} \,. 
\end{eqnarray}


The EOS for matter with hadrons is constructed with the MS model, and
we considered models both with and without hyperons.  In addition, we
considered models incorporating a range of parameters $\zeta$ and
$\xi$.  Two quark Lagrangians were selected, the NJL model with
parameters given by \cite{Rehberg} and the MIT bag model with $150\le
B$/(MeV fm$^{-3}$)$\le250$. 

The choice $\zeta=\xi=0$ maximizes the quark content of matter at a
given density, since the hadronic EOS is stiffest for this case.  This
is illustrated in the left panels of Figure~\ref{chinjl}, which shows the
hadron volume fraction $\chi$ as a function of density for three
representative hadronic parameter sets (neglecting hyperons) for cold
matter without neutrinos.  The NJL (MIT) quark model is shown in the
upper (lower) panel.  However, for a given stellar mass, the quark
content of a neutron star is actually maximized for the softest
parameter set $\zeta=0.06, \xi=1.5$, as shown in the right panels of
Figure~\ref{chinjl}.  This counterintuitive behavior occurs because the
central densities achieved for a given stellar mass are greater for a
softer EOS.  Note that the maximum mass decreases with increasing
softness of the hadronic EOS, which is as expected.

A more intuitive behavior results from variations in the parameters of
the quark Lagrangian, which are explored in Figure~\ref{chinjl}.  The
parameters of the NJL model are relatively well constrained by
experiment.  However, the MIT bag model parameter $B$ is only
constrained by the requirement that the quark-hadron transition not
occur too close to $n_0$, which implies that $B$ is larger than about
125-150 MeV fm$^{-3}$.  The hadron volume fraction is displayed for
the same hadronic parameters, but for different values of $B$, for the
MIT bag model in Figure~\ref{chimit}.  The left panels show variations
with density and the right panels show variations with stellar mass.
The upper panels neglect hyperons while the lower panels include them.
Smaller values for $B$ result in a larger quark content at a given
density, a larger quark content for a given stellar mass, and a
smaller maximum mass.  Note that there is little qualitative change
produced by including hyperons.

In the remainder of this paper, we choose $\zeta=\xi=0$ for the
hadronic parameters and $B=200$ MeV fm$^{-3}$ for the MIT bag
constant.  Figures \ref{chinjl} and \ref{chimit} illustrate the
qualititative changes in quark composition induced by parameter
variations.  It is clear that the mixed phase of quarks and hadrons
can exist in neutron stars at least in the range of 1.2--2 M$_\odot$,
depending on the model.  Ref. \cite{Schaffner} concluded that the
mixed phase is unlikely to exist in neutron stars with masses around
1.4 M$_\odot$ neutron stars, using the NJL Lagrangian.  However, this result
appears to be dependent upon the hadronic interactions.

In the remainder of this paper, we shall consider in detail four EOSs:
hadrons with and without hyperons for the NJL and MIT quark models.

The pressure of matter as a function of the density in units of $n_0$,
$u=n_B/n_0$, is shown in Figure~\ref{pres} for these four cases.  The
mixed phase, indicated by thick lines, is marked by a pronounced
softening of the EOS, observable as a large decrease in
the incompressibility $\partial P/\partial n$ of matter.  The
introduction of hyperons, or a large trapped neutrino fraction,
suppresses the appearance of quarks for both quark models.  The reason
for this is that the additional contribution to the pressure from the
neutrinos or the hyperons is more than cancelled by the addition of a
degree of freedom to the system.  A decrease in the pressure of the
hadronic EOS forces the mixed phase to higher densities,
because the hadronic pressure is not sufficient to match that of the
quark phase until a higher density.  Large amounts of trapped
neutrinos produce a pronounced net increase in the pressure, however,
because the (EOS-softening) transition is shifted to
higher densities in all cases.

It is worth noting that the increase in pressure normally observed for
finite-temperature matter compared to zero-temperature matter
\cite{Pons99,prep} is reversed in the mixed phase produced by quarks.
This reversal does not occur for a mixed phase with kaon condensation
\cite{Pons00}.  This reversal originates in the fact that the phase
transition begins at a lower density at finite temperature, so that
the EOS softens at an earlier density.  Even a small change in the
threshold density of appearance for the mixed phase results in a
significant net decrease of pressure at a fixed density.

The temperature as a function of baryon density for fixed entropy and
net lepton concentration is presented in Figure~\ref{temp}, which
compares the cases ($s=1, Y_{L_e}=0.4$) and ($s=2, Y_{\nu_e}=0$).  In
addition to the cases in which quarks appear, the results ignoring
quarks are also displayed for reference.  The temperature for a
multicomponent system in a pure phase can be analyzed by referring to the
relation for degenerate Fermi particles
\begin{eqnarray}
T=\frac{s}{\pi^2}\left(\sum_i{\frac{{\sqrt{ p_{F,i}^2
 + (m_i^*)^2 }}}{p_{F,i}^2}}\right)^{-1}\,,
\end{eqnarray}
where $m^*_i$ and $p_{F_i}$ are the effective mass and the Fermi
momentum of component $i$, respectively.  This formula is quite
accurate since the hadron and quark Fermi energies are large compared
to the temperature.  The introduction of hyperons or quarks lowers the
Fermi energies of the nucleons and simultaneously increases the
specific heat of the matter, simply because there are more components.
In the case of quarks, a further increase, which is just as
significant, occurs due to the fact that quarks are rather more
relativistic than hadrons.  The combined effects for quarks results in
an actual reduction of temperature with increasing density along an
adiabat.  These results are suggestive that the temperature will be
smaller in a PNS containing quarks than in stars without quarks.
The large reduction in temperature might also influence neutrino
opacities, which are generally proportional to $T^2$.  However, a PNS
simulation is necessary to consistently evaluate the thermal
evolution, since the smaller pressure of quark-containing matter would
tend to increase the star's density and would oppose this effect.

The particle concentrations as functions of density are displayed in
Figures~ \ref{comp1} and \ref{comp2} for the four EOSs considered
here.  The major difference that the choice of quark models produces
concerns the concentration of the strange quark.  The strange quark
(dynamical) 
mass in the NJL model is much larger, by a factor of 2 to 3, than that
assumed in the MIT bag model.  This noticeably reduces its chemical
potential, and hence its concentration, in the NJL model.  The
inclusion of hyperons does not produce significant changes to the
nucleon or electron concentrations, although the electron
concentration begins to fall at the threshold density for the
appearance of hyperons which is lower than the low-density boundary of
the mixed phase region.  
The muon concentration is generally much smaller than that of the electron and 
is omitted from the figures for the sake of clarity. 
By moving the mixed phase region to higher
densities, the inclusion of hyperons, somewhat reduces the width of
the mixed phase region.  This is not apparent in the figures for the
MIT bag case, however, because the mixed phase extends to rather large
densities.

The mass-radius trajectories, computed from the standard relativistic
stellar structure equations, are displayed in Figure~\ref{mrad} for
the four EOSs and for the set of three thermodynamic conditions.
Configurations in which the center of the star is in the mixed phase
region are shown as bold lines.  Although the results displayed are
for a single parametrization of the hadronic matter, it is clear that
neutron stars containing a mixed phase have a moderate range of
masses.  This range could be enhanced by altering either the hadronic
or the quark matter EOS.  The range of masses of stars
containing a mixed phase appears to be diminished in the case that
hyperons exist, as noted in Refs. \cite{PCL,Schaffner}, but this result
is somewhat dependent upon the hyperon coupling constants in addition
to the hadronic and quark matter EOSs.

Coinciding with the result in Figure~\ref{pres} that the quark-hadron
transition at finite temperature occurs at a lower density than at
zero temperature, and thereby reduces the pressure in the mixed-phase
region, the neutrino-free stars with $s=2$ have smaller maximum masses
than those for cold $s=0$ stars.  Nevertheless, the pressure in the
range $n_0<n<1.5n_0$ increases with entropy.  This increase in
pressure results in larger stellar radii for stars below the maximum
mass, a result consistent with the general results found by
Ref.~\cite{LP}.  The difference of pressures between the ($s=0,
Y_{L_e}=0.4$) and ($s=1, Y_{\nu_e}$) cases is much smaller, and
produces relatively less of a radius change.

It is immediately apparent that in all cases shown, a range of masses
are metastable, a condition which exists if the initial PNS
configuration has a greater maximum mass than the final configuration
\cite{PCL}.  This result was foreshadowed by the results presented in
Figure~\ref{pres}, in which the pressure for the lepton-rich
configuration was much larger in the mixed phase than for the other
configurations.  In addition, as the neutrinos diffuse from the star,
the mixed phase shifts to lower densities and so a greater proportion
of the center of the star is in the mixed phase.  In the cases shown,
the maximum mass occurs when the star's central density is in the
mixed phase region.  In other words, pure quark configurations seem
unlikely to occur.

This last point is highlighted in Figure~\ref{phase} which shows phase
diagrams for the mixed phase in the baryon density-neutrino fraction
plane.  The upper and lower boundaries of the mixed phase region are
displayed as bold lines, while the central densities of the maximum
mass configurations are shown as light lines.  In no case, for either
quark model and whether or not hyperons are included, are pure quark stars
possible.  The high-density phase boundaries are always well
above the central densities.


In summary, it is possible for a mixed phase to exist in a neutron
star of virtually any mass above 1.4 M$_\odot$.  Depending upon the
EOSs, a mixed phase is more likely to exist in stars larger than $1.5$
\msun. The precise stellar mass above which a mixed phase containing
quarks might exist depends on the ``softness'' of the hadronic EOS and
the effective bag pressure of the quark model.  Although the quark
content of matter at a given {\it density} is maximized for stiffer
hadronic equations of state, the extent of the mixed phase region in a
neutron star of a given mass is maximized for softer hadronic EOSs.
We have shown that only in extreme cases could a pure quark star
result.

This mixed phase is delayed until most neutrinos have diffused from
the star, leading to the possible metastability of PNSs,
a robust result which depends only on the existence of quarks in dense 
matter.  Finite
temperature permits the quark-hadron transition to occur at slightly
lower densities than at zero temperature, but in a newly-formed PNS
this effect is swamped by the large trapped neutrino fraction which
has the opposite tendency.  Furthermore, $\partial T/\partial n<0$
along adiabats in the quark-hadron mixed phase, a behavior opposite to
that generally found in a mixed phase region containing a kaon
condensate.  This implies that core temperatures may be significantly
lower in stars containing quarks than in those not containing quarks.
Neutrino opacities, which are sensitive to temperature, will be
affected, but the implications for the emitted neutrino fluxes and
temperatures can only be reliably evaluated in the context of a full
PNS simulation.


We acknowledge the support
of the U.S. Department of Energy under contracts
DOE/DE-FG02-88ER-40388 (AS and MP) and
DOE/DE-FG02-87ER-40317 (JML).

\bibliography{spl}
\bibliographystyle{unsrt}

\pagebreak


\section*{Figure Captions}
\begin{figure}
\begin{center}
\leavevmode
\setlength\epsfxsize{6.0in}
\setlength\epsfysize{7.0in}
\epsfbox{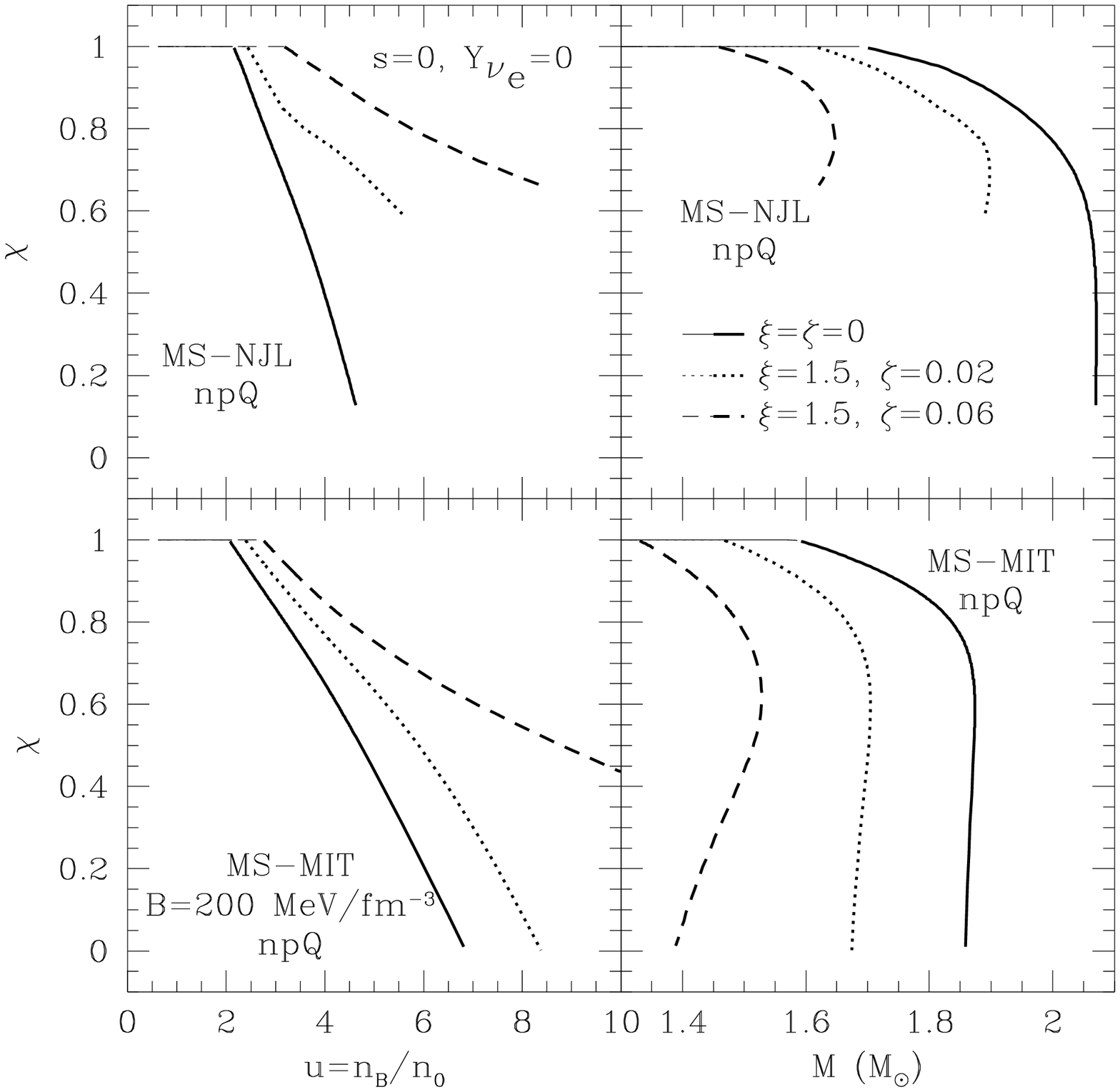}
\caption[]{}
\label{chinjl}
\end{center}
\end{figure}

\ni Figure~\ref{chinjl}: Left panels: The volume fraction of hadrons
as a function of density in units of $n_0$ within the quark-hadron
mixed phase for cold, catalyzed matter ($s=0, Y_{\nu_e}=0$) without
hyperons (npQ).  Three choices for the parameters $\zeta$ and
$\xi$ in the M\"uller-Serot (MS) hadronic Lagrangian 
are illustrated, and the upper panel refers to the Nambu
Jones-Lasinio (NJL) model and the lower panel to the MIT bag model
with $B=200$ MeV fm$^{-3}$.  Right panels: The volume fraction of
hadrons in the star's center as a function of stellar mass for the
same configurations and quark models. \\

\begin{figure}
\begin{center}
\leavevmode
\setlength\epsfxsize{6.0in}
\setlength\epsfysize{7.0in}
\epsfbox{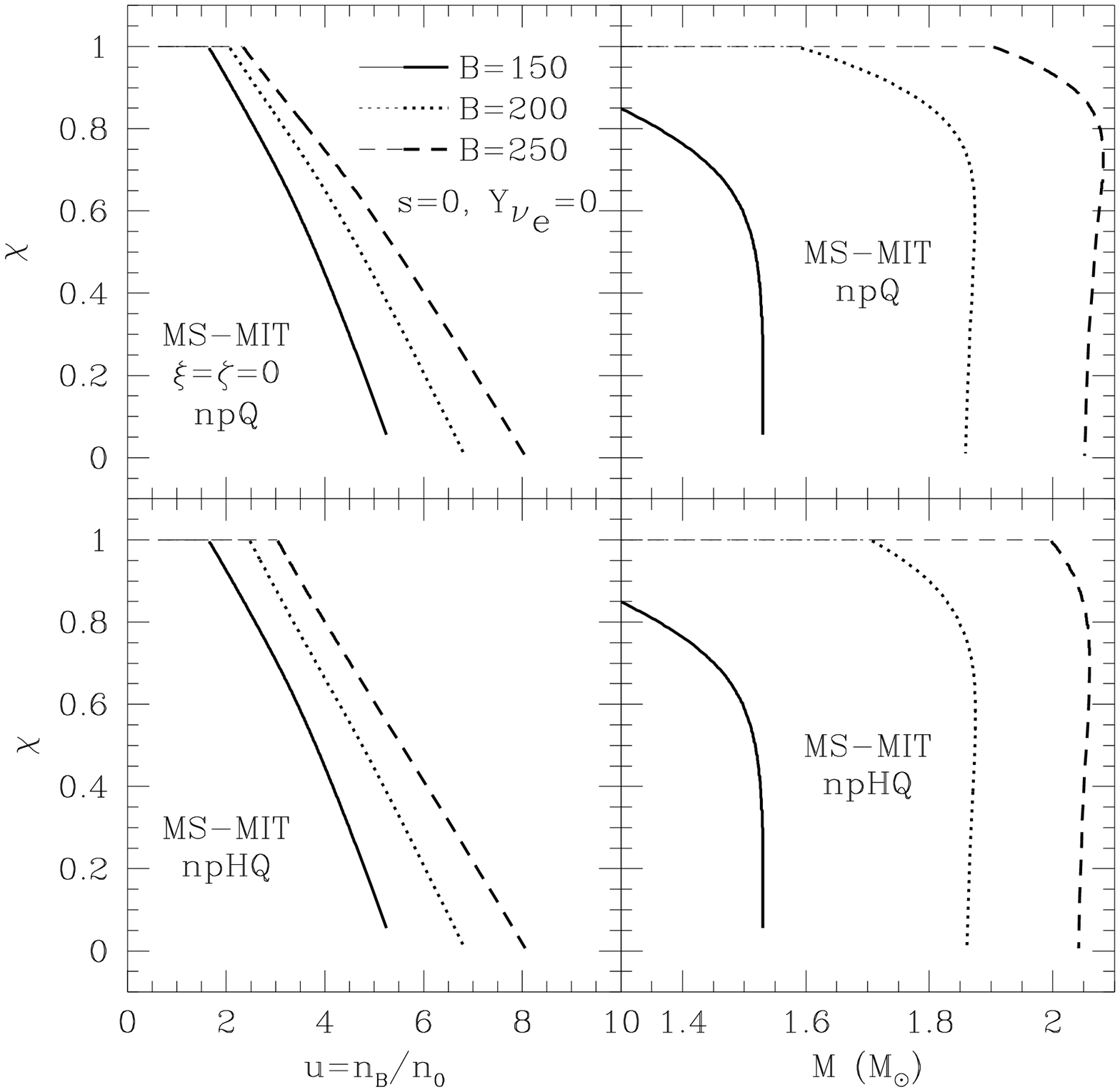}
\caption[]{}
\label{chimit}
\end{center}
\end{figure}

\ni Figure~\ref{chimit}: The same as Figure 1, except that results are
compared for three choices of the bag constant $B$ (in units of MeV
fm$^{-3}$) in the MIT bag model.  Hyperons are ignored in the top
panels (npQ) and included in the bottom panels (npHQ).  The 
parameters $\zeta=\xi=0$ in the M\"uller-Serot (MS) hadronic 
Lagrangian are chosen.\\

\begin{figure}
\begin{center}
\leavevmode
\setlength\epsfxsize{6.0in}
\setlength\epsfysize{7.0in}
\epsfbox{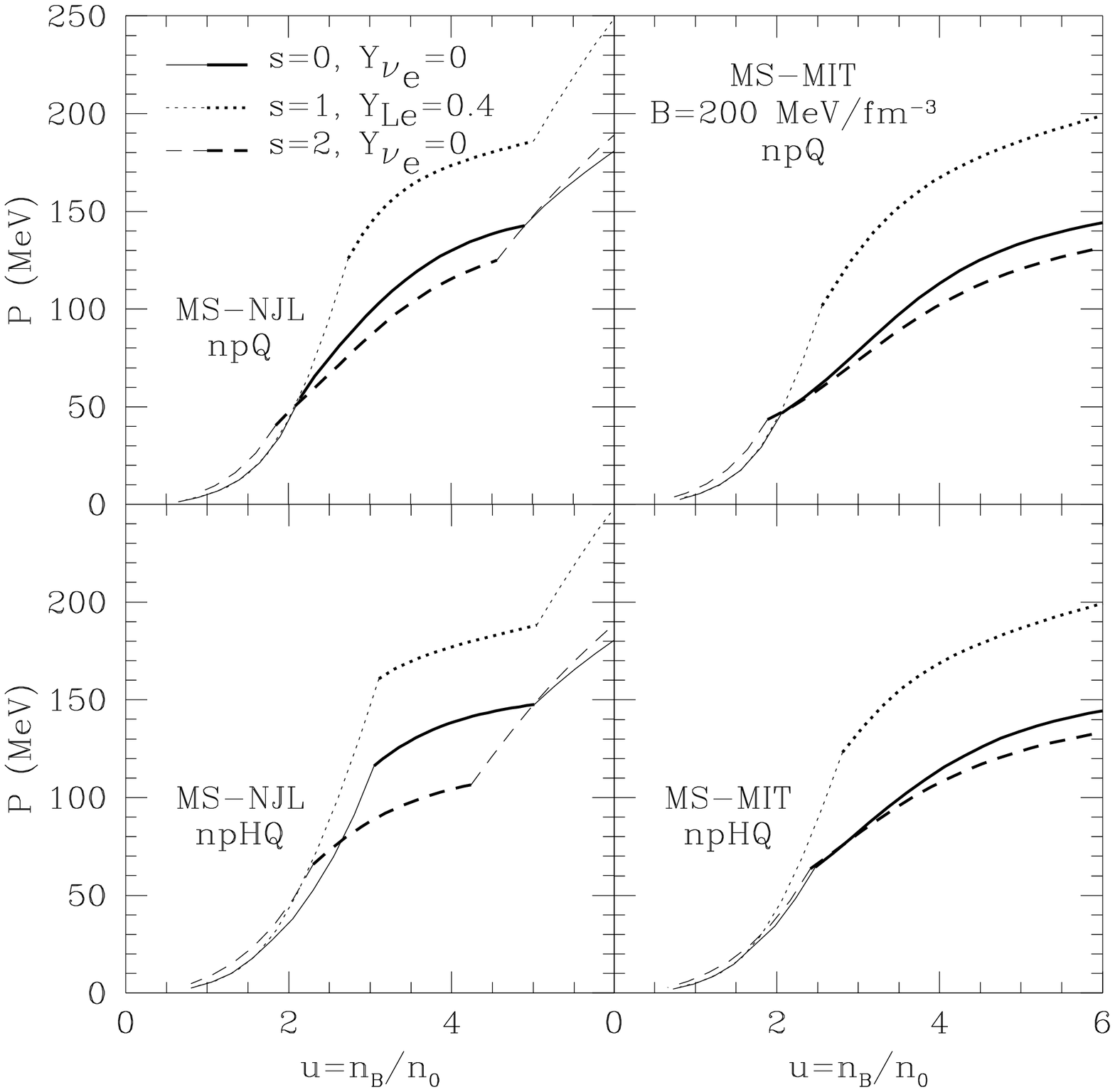}
\caption[]{}
\label{pres}
\end{center}
\end{figure}

\ni Figure~\ref{pres}: Pressure versus density in units of $n_0$ for
three representative snapshots during the evolution of a proto-neutron
star. The top (bottom) panels display results without (with) hyperons,
and the left (right) panels utilize the NJL (MIT bag) quark EOS. 
The parameters $\zeta=\xi=0$ in the M\"uller-Serot (MS) hadronic 
Lagrangian are chosen.
Bold curves indicate the mixed phase region.\\

\begin{figure}
\begin{center}
\leavevmode
\setlength\epsfxsize{6.0in}
\setlength\epsfysize{7.0in}
\epsfbox{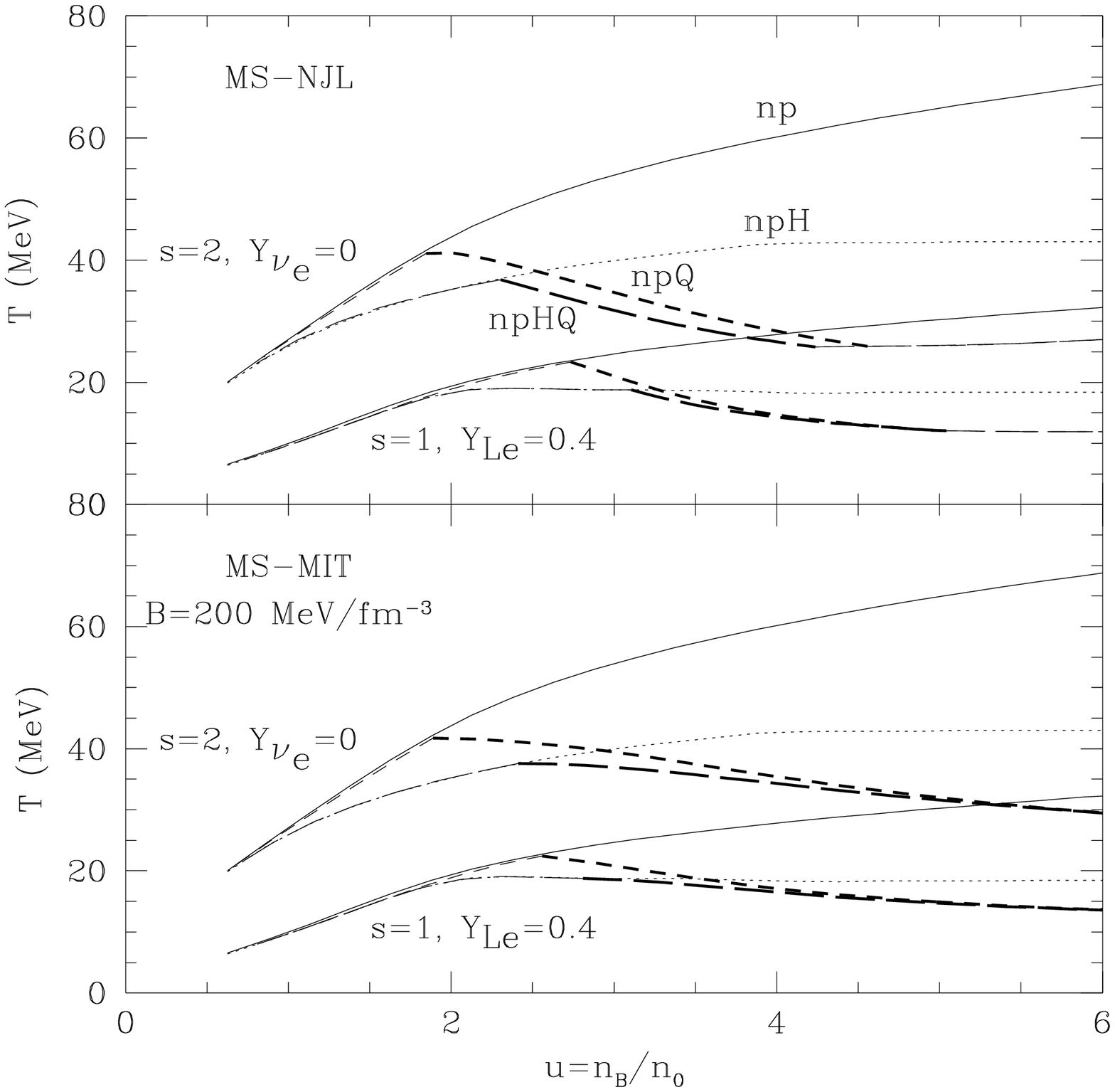}
\caption[]{}
\label{temp}
\end{center}
\end{figure}

\ni Figure~\ref{temp}: Temperature versus density in units of $n_0$
for two PNS evolutionary snapshots. The upper (lower) panel displays
results for the NJL (MIT bag) Lagrangian.  
The parameters $\zeta=\xi=0$ in the M\"uller-Serot (MS) 
hadronic Lagrangian are chosen.  Results are compared for
matter containing only nucleons (np), nucleons plus hyperons (npH),
nucleons plus quarks (npQ) and nucleons, hyperons and quarks (npHQ).
Bold curves indicate the mixed phase region.  \\

\begin{figure}
\begin{center}
\leavevmode
\setlength\epsfxsize{6.0in}
\setlength\epsfysize{7.0in}
\epsfbox{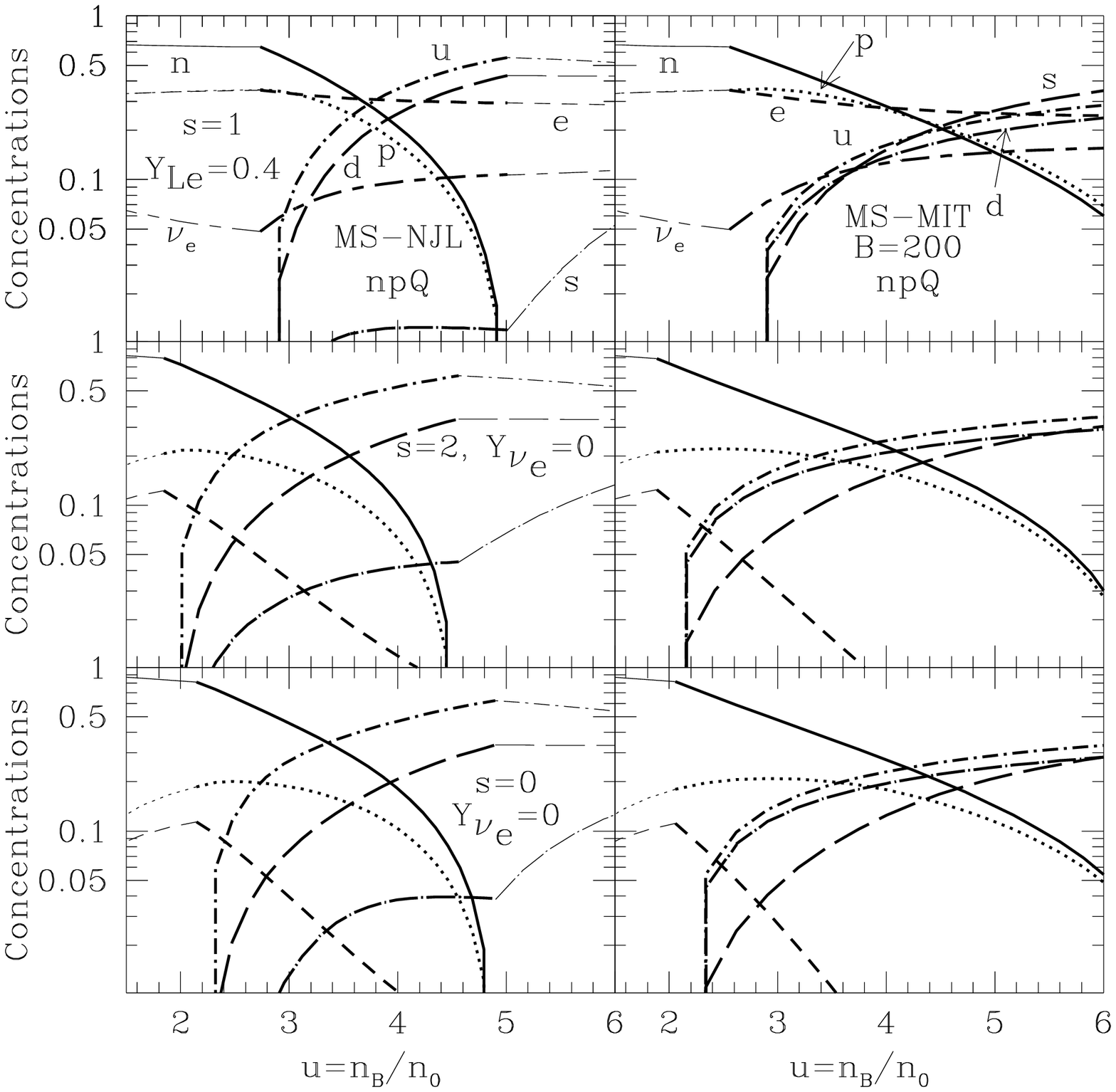}
\caption[]{}
\label{comp1}
\end{center}
\end{figure}

\ni Figure~\ref{comp1}: The concentrations of hadrons, quarks, and
leptons as functions of density in units of $n_0$.  Three
representative snapshots during the evolution of a proto-neutron star
are displayed.  Matter is assumed to contain nucleons and quarks (npQ). 
The parameters $\zeta=\xi=0$ in the M\"uller-Serot (MS) 
hadronic Lagrangian are chosen.
Bold curves indicate the mixed phase region.\\

\begin{figure}
\begin{center}
\leavevmode
\setlength\epsfxsize{6.0in}
\setlength\epsfysize{7.0in}
\epsfbox{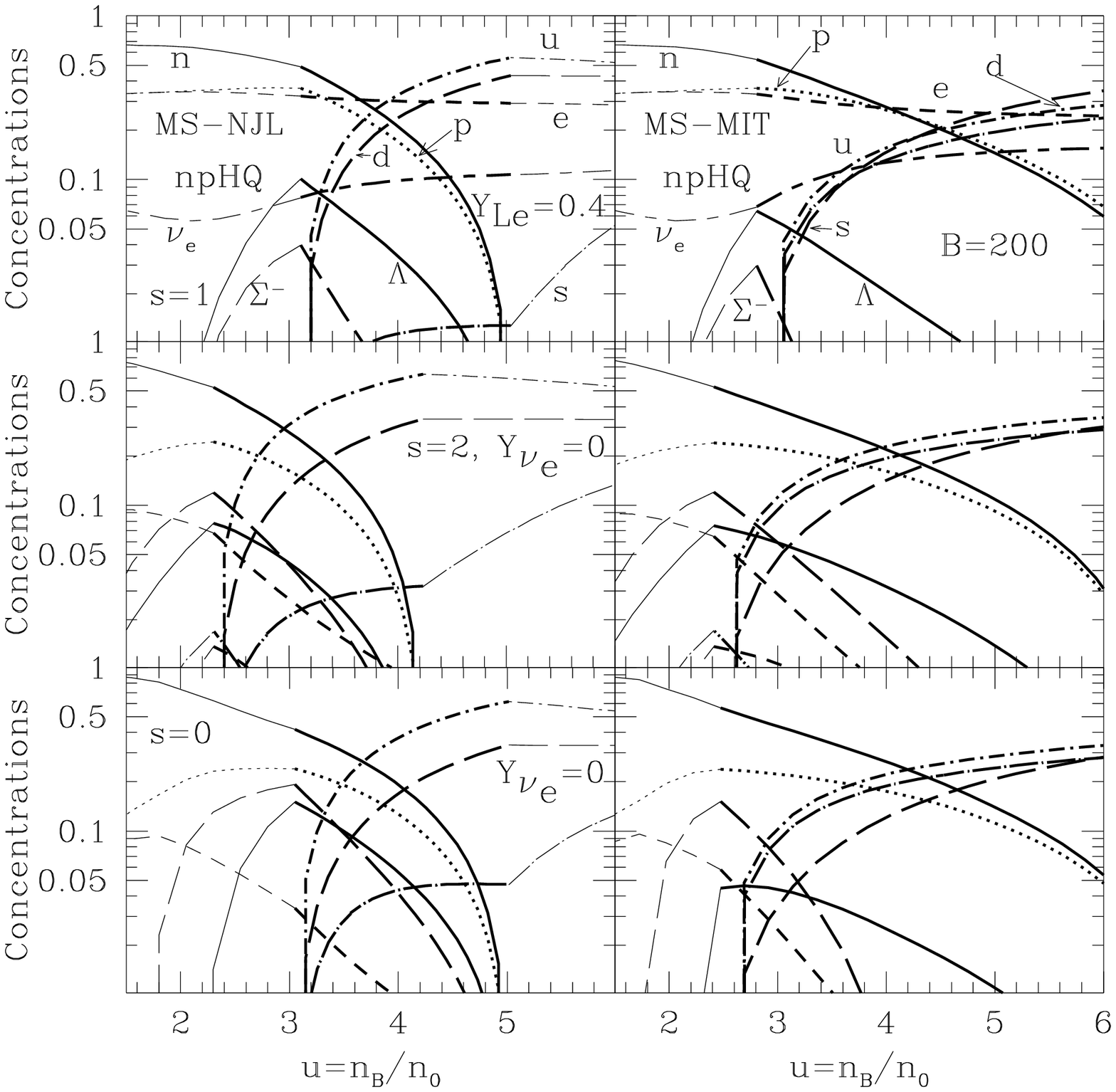}
\caption[]{}
\label{comp2}
\end{center}
\end{figure}

\ni Figure~\ref{comp2}: The same as Figure~\ref{comp1}, except that hyperons
are included (npHQ). \\

\begin{figure}
\begin{center}
\leavevmode
\setlength\epsfxsize{6.0in}
\setlength\epsfysize{7.0in}
\epsfbox{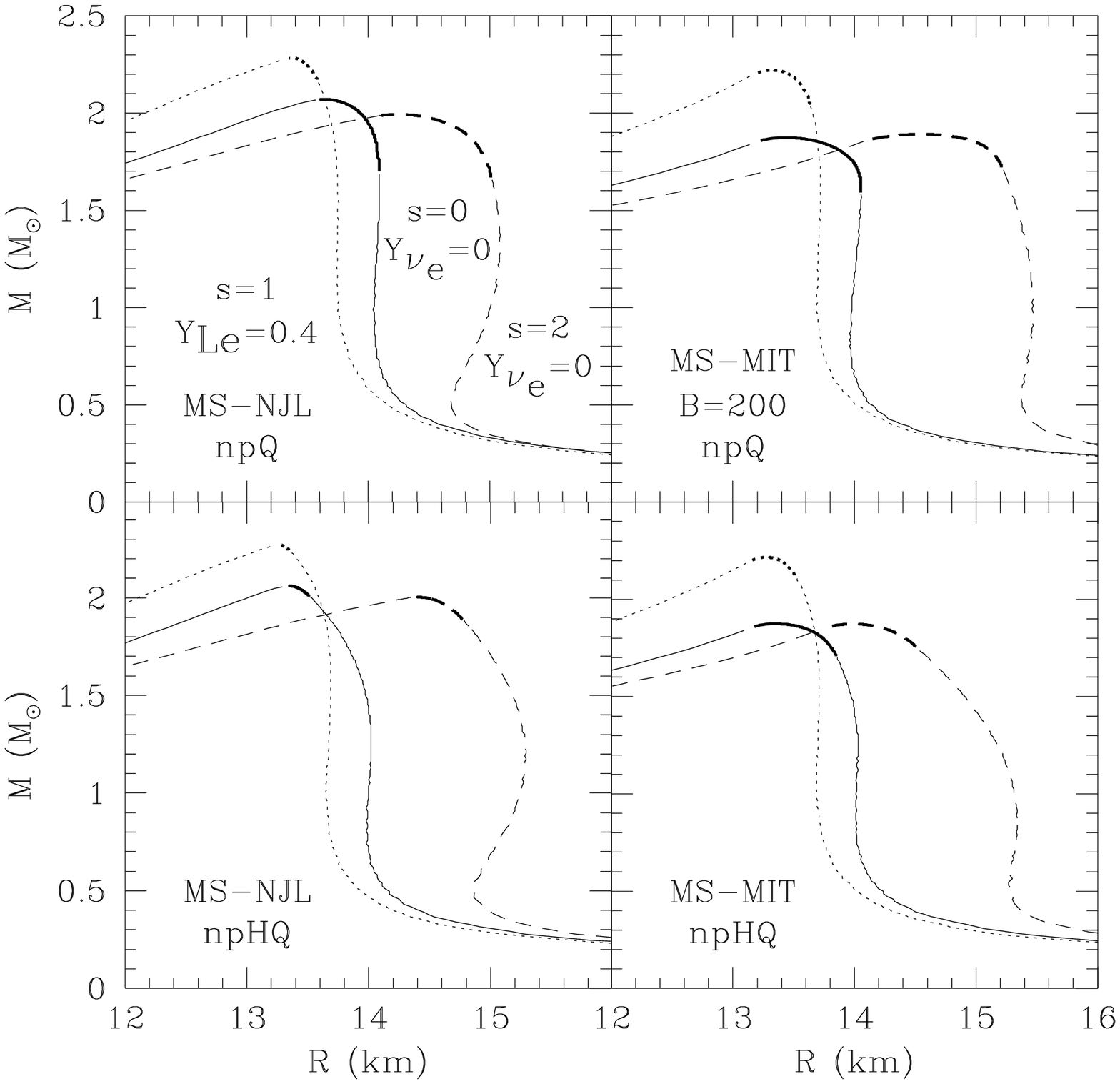}
\caption[]{}
\label{mrad}
\end{center}
\end{figure}

\ni Figure~\ref{mrad}: The gravitational mass versus radius, for three
representative snapshots during the PNS evolution.  The left (right)
panels are for the NJL (MIT bag) quark EOS, and hyperons are (are not)
included in the bottom (top) panels.  
The parameters $\zeta=\xi=0$ in the M\"uller-Serot (MS) 
hadronic Lagrangian are chosen. Bold lines indicate
configurations with a mixed phase at the star's center.\\

\begin{figure}
\begin{center}
\leavevmode
\setlength\epsfxsize{6.0in}
\setlength\epsfysize{7.0in}
\epsfbox{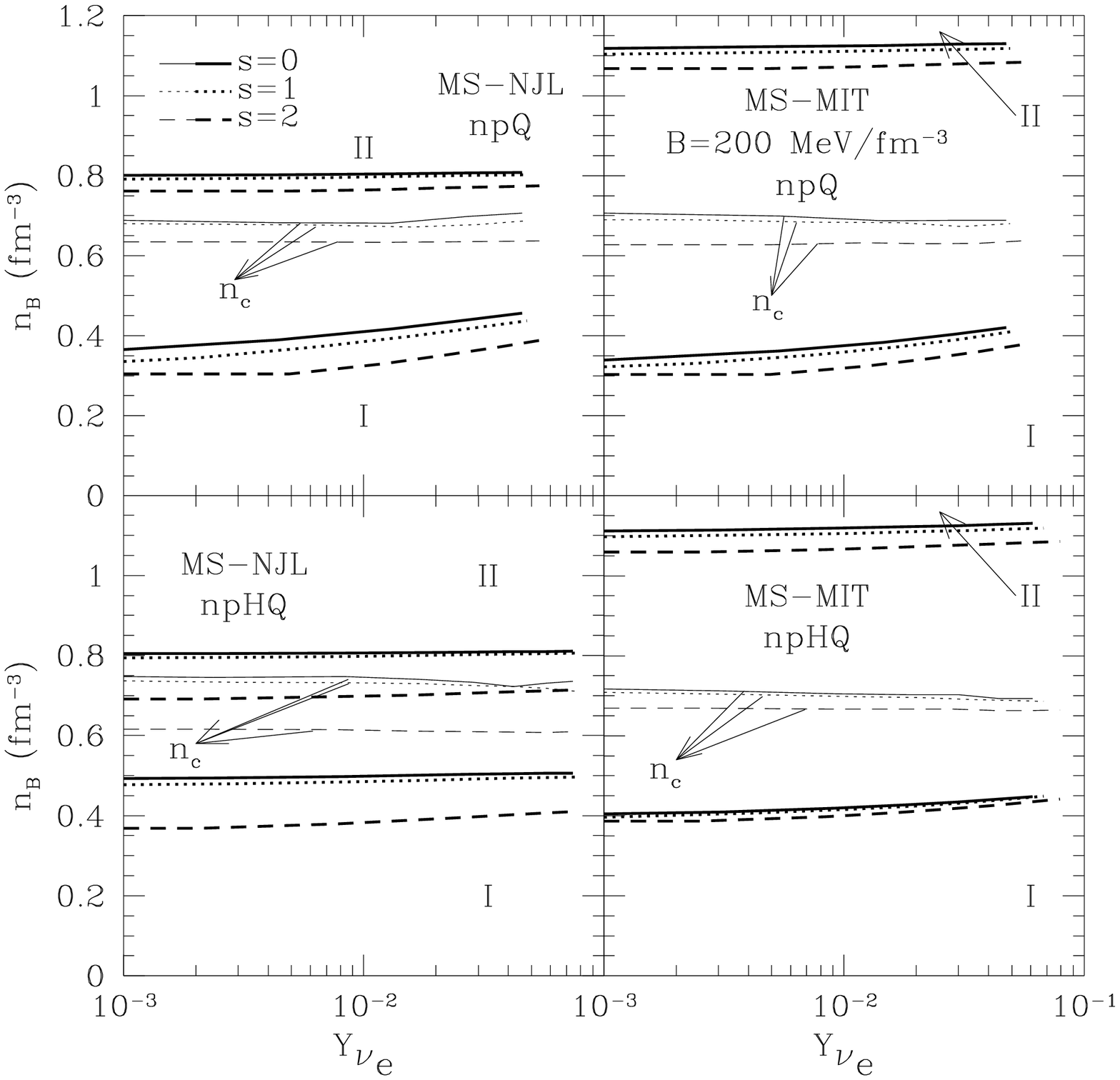}
\caption[]{}
\label{phase}
\end{center}
\end{figure}

\ni Figure~\ref{phase}: The phase diagram of the quark-hadron
transition in the baryon number density - neutrino concentration plane
for three representative snapshots during the evolution of a
proto-neutron star.  The left (right) panels are for the NJL (MIT bag)
quark EOS, and hyperons are (are not) included in the bottom (top)
panels.  
The parameters $\zeta=\xi=0$ in the M\"uller-Serot (MS) 
hadronic Lagrangian are chosen. 
The lower- and upper-density boundaries of the mixed phase
are indicated by bold curves.  The central densities of maximum mass
configurations are shown by thin curves.

\newpage

\newpage

\newpage

\newpage

\newpage

\newpage

\newpage

\end{document}